\documentclass[10pt]{iopart}

\usepackage{graphicx}

\begin{document}

\title[Noise reduction in interferometers]
{Noise reduction in gravitational wave interferometers using feedback}

\author{David Vitali$^{1,2}$, Michele Punturo$^{2}$,
Stefano Mancini$^{1,2}$, Paolo Amico$^{2}$, Paolo Tombesi$^{1,2}$
}
\address{$^{1}$INFM, Dipartimento di Fisica,
Universit\`a di Camerino,
I-62032 Camerino, Italy
}
\address{$^{2}$INFN, Sezione di Perugia,
Perugia, Italy}

\begin{abstract}
We show that the quantum locking scheme recently proposed by Courty {\it et al.} [Phys. Rev. Lett. {\bf 90}, 083601 (2003)]
for the reduction of back action noise
is able to significantly improve the sensitivity of the next generation of gravitational wave interferometers.
\end{abstract}



\maketitle

\section{Introduction}

Radiation pressure noise plays a negligible role in current gravitational wave (GW) detectors.
For example, the VIRGO detector expected sensitivity at low frequency is dominated by the thermal noise of the suspensions and of the test masses \cite{senscurve}.
The limited sensitivity of current GW detectors is reducing the detection distance, for signals like the coalescing binaries, at few MParsec
with a very low expected detection rate. For this reason, several activities have been started in order to improve the current detector performance.
The low frequency sensitivity ($10 - 200$~Hz) is expected to be improved by reducing the thermal noise of the suspension through the adoption
of a monolithic design of the
suspension \cite{FSPerugia}, as already adopted in the GEO600 detector \cite{GEO}, and/or cooling down the temperature to few kelvin degrees.
The high frequency sensitivity ($200 - 10^4$~Hz) is expected to be improved through shot noise suppression,
achieved by increasing the power circulating in the Fabry-Perot cavities of the detector. This is obtained by
increasing the cavity finesse and the laser power up to
few hundreds of watts. The combined effect of these procedures however promotes the radiation pressure noise in the interferometer
to a dominant role at low frequency. Recently, Courty {\it et al.} have proposed a ``quantum locking'' scheme \cite{qlock1}, i.e.,
the use of an active control which is able, in principle, to completely suppress
the radiation pressure noise \cite{qlock2}. Active controls have been already proposed \cite{MVTPRL} and used \cite{HEIPRL,PINARD,amico} for thermal noise suppression
in Fabry-Perot cavities with a movable mirror. Moreover, it has been shown that feedback schemes can be fruitfully applied in a deep quantum regime
\cite{courty,PRALONG} and that they can reduce the quantum back-action noise of a measurement \cite{howmanc},
i.e., the noise on the observable
canonically conjugate to the explicitly measured one. GW interferometers
are based on phase-sensitive measurements and in this case the quantum back-action is light intensity noise, i.e., radiation pressure noise.
The analysis of Refs.~\cite{courty,PRALONG,howmanc} therefore suggests that a feedback control can be
extremely useful for reducing the radiation pressure noise. Refs.~\cite{qlock1,qlock2} considered only shot noise and radiation pressure noise
in their analysis. Here we shall evaluate the capabilities of the quantum locking idea in the presence of additional noises such as thermal and seismic
noise, taking also into account the effects of inefficient detection. In particular we shall apply
the feedback scheme to the case of a realistic gravitational wave interferometer.

\section{The double cavity system}

The idea of quantum locking consists in applying a feedback loop able to lock the motion of an end cavity mirror
to that of a less noisy and controllable reference mirror \cite{qlock1,qlock2}. Under ideal conditions,
the radiation pressure, thermal and seismic noise acting on the mirror can be suppressed and
replaced by the corresponding, smaller noises
of the control mirror, while the shot noise contribution is unaffected by quantum locking.

\begin{figure}
\begin{center}
\includegraphics[width=0.7\textwidth]{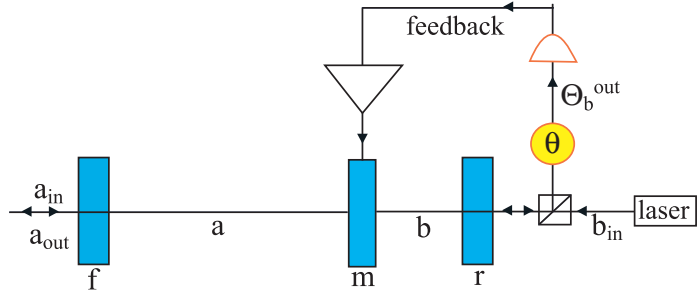}
\end{center}
\vspace{-0.5cm} \caption{\label{fig1}
Schematic description of the feedback scheme. The longer cavity with mode $a$ is one arm of a GW interferometer,
and the shorter cavity with mode $b$ is a control cavity, whose output is measured and used for a phase sensitive feedback loop.
The block marked with $\theta$ denotes a generic phase sensitive detection apparatus
for the measurement of the field quadrature $\Theta_b^{out}$ with adjustable phase $\theta$ (see Section 3).
}
\end{figure}

The experimental scheme is described in Fig.~1, where mirror $f$ and mirror $m$ represent two cavity mirrors of one arm of a GW interferometer
and mirror $r$ is a an additional mirror, forming together with $m$ a shorter control cavity which has the purpose of reducing the noise acting on
mirror $m$. Each cavity is driven by an intense laser beam so that the whole system reaches a steady state in which each internal mode
of the two cavities (with annihilation operator $a$ and $b$ respectively) is in an intense coherent state
(with amplitude $\alpha_a$ and $\alpha_b$ respectively) and the mirror equilibrium positions
are displaced with respect to those in the absence of light by the radiation pressure of the intracavity field.
We are interested in the dynamics of the fluctuations of the mirror and of the cavity field quadratures around this steady state.
Their interaction is described by the effective Hamiltonian \cite{HAMI}
\begin{equation}\label{hint}
  H_{int}=-\frac{2 \hbar \omega_a \alpha_a}{L_a}X_a(Q_m-Q_f)-\frac{2 \hbar \omega_b \alpha_b}{L_b}X_b(Q_r-Q_m),
\end{equation}
where $Q_j$ ($j=m,r,f$) are the mirror displacement operators, $X_l=(l+l^{\dagger})/2$ ($l=a,b$)
are the field amplitude quadrature fluctuations, $L_l$ are the equilibrium cavity lengths and $\omega_l$ the corresponding mode
angular frequencies. Both cavities are usually operated at resonance, i.e., the driving laser frequency is tuned in order to
cancel the cavity frequency shift caused by the radiation pressure.
At resonance, the amplitude quadrature $X_l$ is decoupled from the mirror motion and from the
phase quadrature $Y_l=
(l-l^{\dagger} )/2i$, which is instead affected by the mirror motion. In fact, using Eq.~(\ref{hint}) and passing to the frequency domain
one has
\begin{eqnarray}\label{modes1}
  X_l(\omega)& = & \frac{\sqrt{\gamma_l}}{\gamma_l+2i\omega}X_l^{in}(\omega) \;\;\;\;l=a,b \\
  Y_a(\omega)& = & \frac{\sqrt{\gamma_a}}{\gamma_a+2i\omega}Y_a^{in}(\omega) +\frac{\xi_a (\omega)}{\sqrt{\gamma_a}} \left[Q_m(\omega)
  -Q_f(\omega)\right]
  \label{modes2} \\
  Y_b(\omega)& = & \frac{\sqrt{\gamma_b}}{\gamma_b+2i\omega}Y_b^{in}(\omega) +\frac{\xi_b (\omega)}{\sqrt{\gamma_b}}
 \left[Q_r(\omega)-Q_m(\omega)\right], \label{modes3}
\end{eqnarray}
where $\gamma_l$ is the decay rate of mode $l$ and
$X_l^{in}(\omega)$ and $Y_l^{in}(\omega)$ are respectively the input amplitude and phase noise quadratures of the $l$ mode.
These two noises are uncorrelated
and characterized by the symmetrized correlation functions
\begin{equation}\label{corre}
  \langle X_l^{in}(\omega) X_{l'}^{in}(\omega')\rangle_{sym} = \langle Y_l^{in}(\omega) Y_{l'}^{in}(\omega')\rangle_{sym} = \delta(\omega + \omega ')
  \delta_{ll'}.
\end{equation}
In Eqs.~(\ref{modes2}) and (\ref{modes3}) we have also introduced the frequency dependent optomechanical coupling $\xi_l(\omega)$
taking into account the cavity filtering action,
\begin{equation}\label{csi}
  \xi_l(\omega)=  \frac{2 \omega_l \alpha_l}{L_l\sqrt{\gamma_l}}\frac{1}{1+2i\omega/\gamma_l}=
  \frac{8{\cal F}_l}{1+2i\omega/\gamma_l}\sqrt{\frac{P_l }{2\pi \hbar \lambda_l c}} \;\;\;\; l=a,b,
\end{equation}
where $P_l$ is the input laser power and ${\cal F}_l=\pi c/L_l \gamma_l$ is the finesse of the cavity with mode $l$.
We have also assumed that the central mirror $m$ is perfectly reflecting on both sides, so that each cavity has only one input mirror.

In order to better describe the action of the feedback loop, we neglect the motion of the input mirror $f$, i.e., we assume, for the
moment, $Q_f \simeq 0$. This happens for example when mirror $f$ has a very large mass.
The Fourier transform of the position operator of the two movable mirrors $m$ and $r$ is given by
\begin{eqnarray}\label{foumir1}
  Q_m(\omega) &=& \chi_m(\omega)\left[B_a(\omega)-B_b(\omega)+W_m(\omega)+s_m(\omega) \right]+h(\omega)\frac{L_a}{2} \\
  Q_r(\omega) &=& \chi_r(\omega)\left[B_b(\omega)+W_r(\omega)+s_r(\omega)\right]+h(\omega)\frac{L_a}{2}(1+\epsilon), \label{foumir2}
\end{eqnarray}
where $\chi_j(\omega)$ ($j=m,r$) are the two mirror susceptibilities and
$W_j(\omega)$ and $s_j(\omega)$ ($j=m,r$) are respectively
the thermal noise due to
quantum Brownian motion and the seismic noise acting on mirror $j$. The displacement of mirror $m$ due to a GW is
$h(\omega)L_a/2$, while that of mirror $r$ is $(1+\epsilon)h(\omega)L_a/2$, where
$\epsilon = L_b/L_a$.
The term $B_l(\omega)$ ($l=a,b$) is the Fourier transform of the radiation pressure force noise due to cavity mode $l$, which
is due to the field amplitude quadrature fluctuations $X_l(\omega)$.
Using the optomechanical interaction of Eq.~(\ref{hint}), Eqs.~(\ref{modes1}) and (\ref{csi}), one has
\begin{equation}\label{radpress}
  B_l(\omega) = \hbar \xi_l(\omega) X_l^{in}(\omega).
\end{equation}

\section{The feedback loop}

We now add a feedback loop in order to control the motion of the mirror $m$ of the interferometer. As shown by Eq.~(\ref{modes3}),
the phase quadrature of mode $b$ gives an estimate of the relative mirror distance $Q_m(\omega)-Q_r(\omega)$. The corresponding output signal
can be used to design an active control able to stabilize this relative motion (see Fig.~1). This stabilization is equivalent to lock the motion
of the interferometer mirror $m$ to that of the control mirror $r$ \cite{qlock1}. This means that the
effective noise felt by the mirror $m$ becomes that of mirror $r$, which can be locally controlled and minimized without acting directly on the interferometer.
For example, the radiation pressure noise acting on $r$ is due only to mode $b$, and it can be much smaller than that in the
interferometer by using much less optical power in the control cavity.

The detected output field $b^{out}(\omega)$ is determined by the usual input-output relation~\cite{milwal}
\begin{equation}\label{inout}
  b^{out}(\omega)= \eta_b \sqrt{{\gamma_b}} b(\omega)-\eta_b b^{in}(\omega)-\sqrt{(1-\eta_b)\eta_b}b'(\omega),
\end{equation}
where we have considered the effect of a non-unit detection efficiency by modelling a detector with quantum efficiency $\eta_b$
with an ideal detector preceded by a beam splitter with transmissivity $\sqrt{\eta_b}$, mixing the incident field with an uncorrelated vacuum field
$b'(\omega)$ \cite{gard}.

Let us consider a phase-sensitive detection of a generic output quadrature $\Theta_b^{out}(\omega)=
b^{out}(\omega)e^{-i\theta}+b^{out \dagger}(\omega)e^{i\theta} = X_b^{out}(\omega)\cos\theta + Y_b^{out}(\omega)\sin\theta$.
The following treatment explicitly refers to a homodyne measurement of a field quadrature operator \cite{homo}, but it could be
easily extended to other phase-sensitive scheme as the phase modulation or Pound-Drever detection scheme \cite{Pound}. The only
difference in such a case would be in the expression of the shot noise power spectrum contribution: that in the Pound-Drever scheme case differs from that
of the homodyne case only by a numerical factor depending on the modulation amplitude \cite{pound2}.
Using the input-output relation (\ref{inout}) and Eqs.~(\ref{modes1}) and (\ref{modes3}), one gets the following relation between
the detected field quadrature and the mirror relative distance
\begin{eqnarray}\label{output}
  && \Theta_b^{out}(\omega)=  2\eta_b \xi_b(\omega) \sin\theta
  \left[Q_r(\omega)-Q_m(\omega)\right] \\
  && + \eta_b \frac{1-2i\omega/\gamma_b}{1+2i\omega/\gamma_b}\Theta_b^{in}(\omega)
  -\sqrt{(1-\eta_b)\eta_b}\Theta_b '(\omega), \nonumber
\end{eqnarray}
where $\Theta_b^{in}(\omega)
= X_b^{in}(\omega)\cos\theta + Y_b^{in}(\omega)\sin\theta$ and $\Theta_b '(\omega)=b'(\omega)e^{-i\theta}+b'^{\dagger}(\omega)e^{i\theta}$.
The output quadrature $\Theta_b^{out}(\omega)$ provides an estimate of the Fourier component of the mirror position quadrature $Q_m(\omega)$, given by
\begin{equation}\label{estim}
  Q_m^{est,b}(\omega)\equiv -\frac{1}{2\eta_b \xi_b(\omega) \sin\theta }\Theta_b^{out}(\omega)
\end{equation}
and this is just the estimate of $Q_m(\omega)$ which is used for the feedback loop. The added active control can be
generally described by the following feedback force
\begin{equation}\label{fb}
  F_{fb}(\omega) = - k(\omega) Q_m^{est,b}(\omega) ,
\end{equation}
which have to inserted into the square brackets in Eq.~(\ref{foumir1}).
When the feedback transfer function $k(\omega)$ is a positive real function, the control loop adds a restoring force
while when it is pure imaginary (with positive imaginary part) it describes an additional viscous force.
Using Eqs.~(\ref{foumir2}), (\ref{output}) and (\ref{estim}) in Eq.~(\ref{fb}) and inserting the latter into Eq.~(\ref{foumir1}), one
can derive the general expression of the spectral component of the position of mirror $m$ in the presence of feedback
\begin{eqnarray}
 &&  Q_m(\omega)=  S(\omega) + \frac{\chi_m (\omega)}{1+ k(\omega) \chi_m (\omega)}\left\{\hbar \xi_a (\omega)X_a^{in}(\omega)+W_m (\omega)
  +s_m(\omega)\right. \nonumber \\
  && \left.+k(\omega) \chi_r (\omega)\left[W_r (\omega)+s_r(\omega)\right]
   -\frac{k(\omega)}{2\xi_b(\omega)}\frac{\sqrt{\eta_b^{-1}-1}}{\sin\theta} \Theta_b'(\omega)\right. \label{qm} \\
  &&\left. + \frac{k(\omega)}{2\xi_b^*(\omega)}Y_b^{in}(\omega)+
  \left[k(\omega)\left(\chi_r(\omega)\hbar \xi_b(\omega)+\frac{\cot\theta}{2\xi_b^*(\omega)}\right)-\hbar\xi_b(\omega)\right]X_b^{in}(\omega)\right\},
  \nonumber
\end{eqnarray}
where
\begin{equation}\label{signal}
  S(\omega)= h(\omega)\frac{L_a}{2}\left(1+\epsilon\frac{k(\omega) \chi_m (\omega)}{1+ k(\omega) \chi_m (\omega)}\right)
\end{equation}
is the total signal felt by mirror $m$ due to the tidal force of the GW, and all the other are noise terms.
However, one does not detect directly $Q_m(\omega)$ but its estimator $Q_m^{est,a}(\omega)$ derived from the measurement of the
output phase quadrature of the interferometer cavity $Y_a^{out}(\omega)$. Applying the input-output relation (\ref{inout}) to mode $a$ and using
Eq.~(\ref{modes2}) one gets
\begin{equation}\label{estim2}
  Q_m^{est,a}(\omega)\equiv \frac{Y_a^{out}(\omega)}{2\eta_a \xi_a(\omega)}= Q_m(\omega)+
  \frac{Y_a^{in}(\omega)}{2\xi_a^*(\omega)}
   -\frac{\sqrt{\eta_a^{-1}-1}}{2\xi_a(\omega)} Y_a'(\omega),
\end{equation}
where $\eta_a$ is the quantum efficiency of the phase detection of mode $a$ and we have again introduced an additional phase noise
$Y_a'(\omega)$ to account for the imperfect detection.
The last two terms on the right hand side of Eq.~(\ref{estim2}) give the shot noise contribution
and have to be added to the other noise terms in Eq.~(\ref{qm}).

Using Eqs.~(\ref{qm}) and (\ref{estim2}), the fact that the various noise sources are independent, and the
correlation functions (\ref{corre}), which also hold for $\Theta_b'(\omega)$ and
$Y_a'(\omega)$, one can obtain the (double-sided) position noise power spectrum $n_{\theta}(\omega)$. In fact, one has
$Q_m^{est,a}(\omega)=S(\omega)+N_{\theta}(\omega)$, with $\langle N_{\theta}(\omega) N_{\theta}(\omega')\rangle = n_{\theta}^2(\omega)\delta(\omega + \omega ')$ and
\begin{eqnarray}
 &&  n_{\theta}^2(\omega)=  \frac{1}{4 |\xi_a(\omega)|^2\eta_a}+
 \left|\frac{\chi_m (\omega)}{1+ k(\omega) \chi_m (\omega)}\right|^2
 \left\{\hbar ^2 |\xi_a (\omega)|^2 +\Phi_{Ts}^m (\omega)\right. \nonumber \\
  && \left.+|k(\omega) \chi_r (\omega)|^2 \Phi_{Ts}^r (\omega)+ \frac{|k(\omega)|^2}{4|\xi_b(\omega)|^2}\left(1+
  \frac{\eta_b^{-1}-1}{\sin^2\theta}\right) \right. \nonumber \\
  &&\left. + \frac{
  \left|k(\omega)\left(2\chi_r(\omega)\hbar |\xi_b(\omega)|^2+\cot\theta\right)-2\hbar|\xi_b(\omega)|^2\right|^2}{4|\xi_b(\omega)|^2}\right\}.
  \label{noise}
\end{eqnarray}
where we have denoted with $\Phi_{Ts}^j(\omega)$ $(j=m,r)$ the total force noise power spectrum due to the thermal and seismic noise acting on mirror $j$
and which is defined by the relation
\begin{eqnarray}\label{corre2}
  \Phi_{Ts}^j(\omega)\delta(\omega + \omega ') &=& \left[\Phi_{T}^j(\omega)+\Phi_{s}^j(\omega)\right]\delta(\omega + \omega ') \\
  &=& \langle W_j(\omega) W_j(\omega')\rangle_{sym} + \langle s(\omega) s(\omega')\rangle_{sym}\;\;\;\;j=m,r. \nonumber
\end{eqnarray}

\section{Feedback loop optimization}

The above expressions for $S(\omega)$ and $n_{\theta}(\omega)$ hold for a general linear feedback acting on $m$ (see Eq.~(\ref{fb})),
based on the detection of a generic field quadrature $\Theta_b^{out}(\omega)$. Now we have to optimize the loop, i.e., to determine
the best choice for the field phase $\theta$ and for the feedback gain function $k(\omega)$ able to maximize the signal-to-noise ratio (SNR)
$R_{\theta}(\omega)=
S(\omega)/n_{\theta}(\omega)$. In principle, this best choice corresponds to the values $\theta_{opt}$ and $k_{opt}(\omega)$
which maximize $R_{\theta}(\omega)$ as a function of $\theta$ and $k(\omega)$. In practice however, tuning simultaneously
the feedback gain phase and modulus and the phase $\theta$, dependently on each other and at each frequency of operation, is a prohibitive task. It is more
convenient to maximize the SNR first with respect to $\theta$ independently from $k(\omega)$, and then maximize
with respect to $k(\omega)$.
The detection phase $\theta$ appears only in the noise spectrum and
only through $\cot\theta$, so that optimizing means minimizing $n_{\theta}^2(\omega)$  with respect to $\cot\theta$.
Moreover, since we ask for an optimal value $\theta_{opt}$ independent from the feedback gain, we minimize only the terms
of $n_{\theta}^2(\omega)$ which are proportional to $|k(\omega)|^2$ (see Eq.~(\ref{noise})). This essentially coincides with the strategy followed by
Courty {\it et al.} in Ref.~\cite{qlock2} where they also show that the optimal detection phase can be obtained for example
by adding a third tunable cavity within the feedback
loop and suitably adjusting its detuning. Our optimization procedure gives
\begin{equation}\label{thopt}
  \cot\theta_{opt}= -2\hbar |\xi_b(\omega)|^2 \chi_r(\omega) \eta_b,
\end{equation}
corresponding to the generalization of the result  of Ref.~\cite{qlock2} in the case of imperfect detection $(\eta_b < 1)$.
It is evident from Eq.~(\ref{thopt}) that this optimization can be achieved if the susceptibility
 $\chi_r(\omega)$ is essentially real within the operation bandwidth. From now on we shall assume that both mirror susceptibilities
are real, which is verified if the detection bandwidth is far away from any mechanical resonance peak of the mirrors.
For example, in the VIRGO interferometer this condition is verified from $10$ Hz to $1$ kHz \cite{VIRGO}, if we neglect the so--called violin modes.
The noise power spectrum at the optimal detection phase is
\begin{eqnarray}
 &&  n_{opt}^2(\omega)\equiv n_{\theta = \theta_{opt}}^2(\omega)=  \frac{1}{4 |\xi_a(\omega)|^2\eta_a}+
 \left|\frac{\chi_m (\omega)}{1+ k(\omega) \chi_m (\omega)}\right|^2 \nonumber \\
 && \times \left\{\hbar ^2 |\xi_a (\omega)|^2 +\Phi_{Ts}^m (\omega)
 +|k(\omega) \chi_r (\omega)|^2 \Phi_{Ts}^r (\omega)+ \frac{|k(\omega)|^2}{4|\xi_b(\omega)|^2\eta_b} \right. \nonumber \\
  &&\left. + \hbar ^2 |\xi_b(\omega)|^2
  \left[\eta_b + (1-\eta_b) |1-k(\omega)\chi_r(\omega)|^2\right]\right\}.
  \label{noiseopt}
\end{eqnarray}
Finally we have to maximize the SNR $R(\omega)=S(\omega)/n_{opt}(\omega)$ with respect to the complex variable
$k(\omega)$. After tedious but straightforward calculations, one can see that in the case of real mirror susceptibilities $\chi_j(\omega)$
$(j=m,r)$, the maximum $R(\omega)$ is achieved for a {\em real} positive feedback gain function $k(\omega)$
(corresponding to a restoring elastic feedback force), which is given by
\begin{equation}\label{kopt}
  k_{opt}(\omega)=\frac{(1+\epsilon)}{\chi_m(\omega)}\frac{n_{eq}^{2,m}(\omega)}{n_{eq}^{2,r}(\omega)},
\end{equation}
where we have introduced the ``equivalent'' noise power spectra of the two mirrors,
\begin{eqnarray}\label{nkopt1}
 n_{eq}^{2,m}(\omega)&=& \chi_m^2(\omega)\left[\hbar^2|\xi_b(\omega)|^2\left(1+\frac{1-\eta_b}{1+\epsilon}\frac{\chi_r(\omega)}{\chi_m(\omega)}\right)
 +\hbar^2|\xi_a(\omega)|^2 \right.
   \nonumber \\
  &&\left. +\Phi_{Ts}^m (\omega)\right]+\frac{\epsilon}{4|\xi_a(\omega)|^2\eta_a (1+\epsilon)}, \\
  n_{eq}^{2,r}(\omega)&=& \chi_r^2(\omega)\left[ \hbar^2|\xi_b(\omega)|^2(1-\eta_b)\left(1+\frac{\chi_m(\omega)(1+\epsilon)}{\chi_r(\omega)}\right)
   \right. \nonumber \\
  &&\left.+\Phi_{Ts}^r (\omega)\right]-\frac{\epsilon}{4|\xi_a(\omega)|^2\eta_a }+
  \frac{1}{4|\xi_b(\omega)|^2\eta_b }. \label{nkopt2}
\end{eqnarray}
The term $n_{eq}^{2,m}(\omega)$ is, except for some factors, the open-loop noise power spectrum of mirror $m$
without the shot noise term ($\epsilon$ is typically very small), and including the back-action noise of mirror $r$.
Instead $n_{eq}^{2,r}(\omega)$ essentially coincides with the open-loop noise
power spectrum of mirror $r$ without the radiation pressure noise contribution ($\eta_b \simeq 1$ typically).
The quantum locking scheme is useful when mirror $r$ is less noisy than mirror $m$ and in such a situation
the optimal feedback gain of Eq.~(\ref{nkopt1}) is typically large because $n_{eq}^{2,m}(\omega)/n_{eq}^{2,r}(\omega) \gg 1$.
This is true in particular when the noise is mainly determined by the radiation pressure and thermal noise contributions. Instead, if
the dominant contribution to the mirror noise is given by the shot noise term,
the optimal feedback gain of Eq.~(\ref{nkopt1}) is much smaller and the quantum locking
scheme is much less effective because the main effect of the loop is only adding the shot noise of the detection of $\Theta_b^{out}$ to the noise of mirror $m$.

A real feedback gain $k(\omega)$ implies using proportional active controls which are however less stable and more difficult to implement than
derivative or integral controls, which imply using a purely imaginary feedback gain \cite{control}. For example, cold damping feedback schemes \cite{coldd},
in which an additional viscous force is added, are based on a derivative control and they have been already
employed \cite{HEIPRL,amico} for the reduction of the thermal noise
of a cavity mirror. It is therefore interesting to see which is the best achievable SNR $R(\omega)$
if we limit ourselves to a pure imaginary feedback gain $k(\omega)$. From Eqs.~(\ref{signal}) and (\ref{noiseopt}) one can see that in this case
the highest $R(\omega)$ is achieved asymptotically in the limit of very large feedback gain $|k(\omega)| \to \infty$.
In this limit, the interferometer behaves as if the end mirror $m$ were replaced by the less noisy control mirror $r$. In fact, the
signal becomes
\begin{equation}\label{signalasy}
  S_{as}(\omega)\simeq h(\omega)\frac{L_a}{2}(1+\epsilon),
\end{equation}
while the asymptotic noise power spectrum becomes
\begin{eqnarray}
n_{as}^{2}(\omega)&\simeq & \chi_r^2(\omega)\left[ \hbar^2|\xi_b(\omega)|^2(1-\eta_b)
  +\Phi_{Ts}^r (\omega)\right]+ \left[4|\xi_b(\omega)|^2\eta_b \right]^{-1} \nonumber \\
  && +\left[4|\xi_a(\omega)|^2\eta_a\right]^{-1}, \label{nasy}
\end{eqnarray}
which is just the noise power spectrum of mirror $r$ (with the radiation pressure noise reduced by a factor $(1-\eta_b)$ thanks to
the optimization of the detection phase $\theta$), supplemented with the shot noise due to the measurement of mode $a$.
Eqs.~(\ref{signalasy}) and (\ref{nasy}) show very well the idea of quantum locking:
the feedback loop is added in order to ``lock'' the interferometer mirror $m$ to the less noisy and better controllable mirror $r$.
This quantum locking is particularly evident in the limit of very large feedback gain: in this limit,
all the noises acting on mirror $m$ except for the shot noise from the detection of mode $a$ (i.e., radiation pressure, thermal and seismic noise)
are suppressed, and the feedback loop replaces them with the noise acting on the control mirror $r$.
At the same time, the signal is not degraded by the presence of feedback.

Employing a purely imaginary feedback gain is not optimal because we have seen that the maximum SNR is achieved
for the real feedback gain of Eq.~(\ref{kopt}). However it is possible to see that it is {\em nearly optimal}, in the sense that the improvement of $R(\omega)$
with respect to the asymptotic values of Eqs.~(\ref{signalasy}) and (\ref{nasy}) is almost negligible
in situations of practical interest. In fact, the relative SNR improvement $\Delta R(\omega)/R(\omega)$
achievable when using the real optimal gain $ k_{opt}(\omega)$ instead of a large imaginary feedback gain can be written as
\begin{equation}\label{improv}
  \frac{\Delta R(\omega)}{R(\omega)}=\frac{n_{eq}^{2,r}(\omega)}{n_{eq}^{2,r}(\omega)+n_{eq}^{2,m}(\omega)(1+\epsilon)^2 }.
\end{equation}
As discussed above, adding the feedback loop is useful when the mirror $r$ is much less noisy than mirror $m$, i.e.,
when $n_{eq}^{2,r}(\omega)/n_{eq}^{2,m}(\omega) \ll 1$ and in this limit the relative improvement of the SNR
is very small.

The results of the present analysis generalize those obtained by Courty {\it et al.} in Refs.~\cite{qlock1,qlock2}, where only the unavoidable
and fundamental noises of quantum origin, i.e., radiation pressure and shot noise have been considered. Here we show that quantum locking can be used
to control not only the radiation pressure, but also thermal and seismic noise. Differently from Refs.~\cite{qlock1,qlock2}, we have optimized the scheme
by maximizing the SNR rather than minimizing the noise. However, in cases of practical interest, when
the noise on mirror $m$ is much larger than that on mirror $r$ and therefore a large feedback gain is needed, the two optimization strategies
practically coincide and the results are qualitatively the same.

\section{Numerical results for the interferometer case}

The preceding sections have shown how to design and optimize a feedback control able to reduce the noise of a cavity mirror.
It is interesting to see the performance of this control scheme in a realistic GW detection
apparatus, as for example the Michelson interferometer scheme with a Fabry-Perot cavity in each arm used in the VIRGO \cite{VIRGO} and LIGO \cite{LIGO}
experiments. In this case, the cavity with end mirrors $m$ and $f$ (which we now assume equal) represents one interferometer arm
and its output $a^{out}$ is mixed at a $50$--$50$ beam splitter with the output from a second, identical, Fabry-Perot cavity in the orthogonal interferometer arm.
One could think to apply the feedback control described above to each of the four interferometer mirrors, and see if the sensitivity of the gravitational
wave detector is improved by the quantum locking scheme (see also the very recent work of Ref.~\cite{heidnew}, where
a single feedback control is proposed to lock the whole cavity and which therefore requires only two control mirrors for the interferometer).
In the simple, fully symmetric, case of four equal, feedback-controlled, interferometer mirrors,
the signal and noise spectrum of the interferometer
can be simply deduced from those evaluated in the preceding Section for a single mirror.

As shown by Eq.~(\ref{hint}), each cavity mode in one arm is coupled to the {\em relative motion} of the two end mirrors.
This implies that the radiation pressure noise acts in a correlated way on both mirrors, and that when
the two mirrors $m$ and $f$ are equal the total radiation pressure noise in one arm is twice that of the single mirror $m$ (i.e., the
noise power spectrum becomes four times larger). The radiation pressure noises in the two interferometer arms are uncorrelated and therefore
they add in quadrature.
Obviously the shot noise contribution of one arm remains the same if also the motion of the input mirror $f$ is considered; moreover
the shot noises of the two arms are normally uncorrelated, and therefore they add in quadrature.
Finally, the four thermal and the seismic noises acting on each mirror are uncorrelated and they add in quadrature.
Using these arguments,
the phase difference at the output of the interferometer can be written as \cite{Saulson}
\begin{eqnarray}
\Delta \phi (\omega) =
{8 {\cal F}_a \over \lambda_a}
\left( 2\Delta L_a (\omega) + \sqrt{\sum{n_x^2}(\omega) + \sum{n_y^2}(\omega) }
\right) = \nonumber \\
{8 {\cal F}_a \over \lambda_a}
\left( h (\omega)L_a + \sqrt{\sum{n_x^2}(\omega) + \sum{n_y^2}(\omega) }  \right),
\label{sfasTot}
\end{eqnarray}
where $\Delta L_a (\omega)$ is the arm length fluctuation (opposite in the two arms) due to the GW space-time strain $h(\omega)$, while
$\sum{n_x^2}(\omega)$
and $\sum{n_y^2}(\omega)$ denote the total displacement noise spectra of the two orthogonal interferometer arms.
If, for a single mirror, we denote with $n_s^2(\omega)$ the displacement noise spectrum of the seismic noise,
with $n_T^2(\omega)$ the corresponding thermal noise contribution, with $n_{rp}^2(\omega)$ the radiation pressure noise contribution,
and with $n_{sh}^2(\omega)$ the shot noise contribution, using the arguments above, we can express the $SNR=1$ amplitude equivalent--length noise as:
\begin{eqnarray}
\Delta L(\omega)=
{1  \over 2}
\sqrt{4 n_s^2(\omega) + 4 n_T^2(\omega) + 2\cdot 4 n_{rp}^2(\omega) + 2 n_{sh}^2 (\omega)} = \nonumber \\
\sqrt{n_s^2(\omega) + n_T^2(\omega) + 2 n_{rp}^2(\omega) + {1 \over 2} n_{sh}^2(\omega) }.
\label{lengthNoise}
\end{eqnarray}
Hence, we can readapt Eq.~(\ref{noiseopt}) to the case of a complete GW interferometer as:
\begin{eqnarray}
  n_{opt}^2(\omega)=
  \frac{1}{8 |\xi_a(\omega)|^2\eta_a}+
 2 \left|\frac{\chi_m (\omega)}{1+ k(\omega) \chi_m (\omega)}\right|^2
 \left\{\hbar ^2 |\xi_a (\omega)|^2\right\} + \nonumber \\
\left|\frac{\chi_m (\omega)}{1+ k(\omega) \chi_m (\omega)}\right|^2
\left\{
\Phi_{Ts}^m (\omega)
 +|k(\omega) \chi_r (\omega)|^2 \Phi_{Ts}^r (\omega)+ \frac{|k(\omega)|^2}{4|\xi_b(\omega)|^2\eta_b} \right. \nonumber \\
  \left. + \hbar ^2 |\xi_b(\omega)|^2
  \left[\eta_b + (1-\eta_b) |1-k(\omega)\chi_r(\omega)|^2\right]\right\}.
  \label{noiseoptITF}
\end{eqnarray}
The first term in Eq.~(\ref{noiseoptITF}) is the shot noise of the interferometer; the second one is the radiation pressure noise of the interferometer,
taking into account the closed--loop susceptibility
${\chi_m (\omega)}/{(1+ k(\omega) \chi_m (\omega))}$
instead of the open--loop one $\chi_m (\omega)$.
The third term adds the thermal and seismic noise due to the main cavity mirrors to the noises coming from the control cavity. The effect of these noises can be seen as increasing the seismic or thermal noises of each main mirror.
\par
\begin{figure}
\begin{center}
\includegraphics[width=0.7\textwidth]{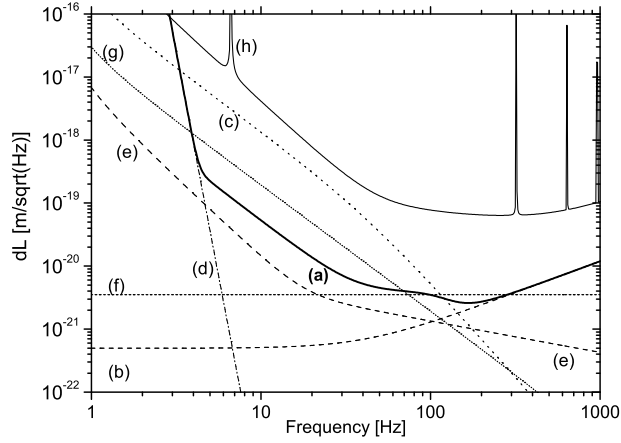}
\end{center}
\vspace{-0.5cm}
\caption{Displacement amplitude spectral density ($m / \sqrt{Hz}$) for the different noise contributions (see Eq.~(\ref{lengthNoise})):
(a) Closed loop sensitivity of the full cryogenic interferometer at the optimal feedback gain, (b) Shot noise of the interferometer,
(c) Open loop radiation pressure noise of the interferometer,
(d) Open loop seismic noise, (e) Open loop thermal noise, (f) Shot noise of the control cavity, (g) Radiation pressure
noise of the control cavity, (h) Expected VIRGO sensitivity \cite{senscurve}.}
\label{fig2}
\end{figure}

We consider the case of an advanced  power--recycled interferometric GW detector, using a seismic filtering system like that of VIRGO \cite{SA1},
and a cryogenic operational temperature to reduce the thermal noise contribution. The susceptibility of a single mirror can be written as
\begin{equation}\label{suscept}
  \chi_m(\omega) = \sum_{j=0}^{Nmodes} \frac{\theta_j /M_j}{\Omega_j^2 -\omega^2 + i \phi_j \Omega_j^2},
\end{equation}
where, for each mode $j$, $\theta_j$ represents the coupling to the horizontal beam direction of the mode displacement, $M_j$ is the equivalent mass, $\Omega_j$ is the resonance frequency and $\phi_j$ is the loss angle.
The displacement (double--sided) power spectrum, due to the thermal noise can be evaluated using the fluctuation--dissipation theorem \cite{FDtheorem}:
\begin{equation}\label{thermaldisplacement}
  n_T^2(\omega)=\left|\chi_m(\omega)\right|^2 \Phi_T^m(\omega) = \hbar
\coth\left(\frac{\hbar \omega}{2 k_B T}\right) \left|{\rm Im}\left\{\chi_m(\omega)\right\} \right|,
\end{equation}
while the displacement power spectrum due to the seismic noise can be modelled as
\begin{equation}\label{seismicdisplacement}
  n_s^2(\omega)=\left|\chi(\omega)_{m}\right|^2 \Phi_s^m(\omega) \simeq \left\{
\left(\frac{\Omega_0}{\omega}\right)^{2 N_s}
\left(\frac{A_0^2}{\omega^2}\right) \right\}^{2},
\end{equation}
where $\Omega_0$ is the pendulum frequency, $N_s$ is the number of pendulum stages used to filter the seismic noise and $A_0$ is a constant that depends on the
local seismic noise amplitude.
In Fig.~\ref{fig2} the displacement amplitude spectral densities obtained with this model are reported. The operational parameter
used in this simulation are reported in Table~\ref{Table:Parameter}. The control mirrors have been chosen equal
to the interferometer mirrors (i.e., equal susceptibilities and temperatures).
The overall gain in sensitivity achieved using the optimal feedback gain is evident. The sensitivity is limited by seismic noise below $5$ Hz and by shot noise
above $200$ Hz, but within the $5 - 200$ Hz frequency interval, the dominant open--loop radiation pressure contribution (curve (c) in Fig.~2)
is suppressed by a factor of $50$ by the closed--loop control, and the sensitivity is two orders of magnitude better than that expected for VIRGO \cite{senscurve}
(curve (h)). The several peaks in the Virgo sensitivity curve are due to the
resonances in the mirror suspension: At about 7 Hz one can see the vertical
resonant mode of the suspension last stage, while between 200 Hz and 1 kHz
the first violin modes are shown. In the calculated plot (curve (a)) these resonances are not reported because, in the
case of a cryogenic interferometer, as we have assumed here, the suspension
design must be completely changed. The vertical mode will surely increase
its frequency while the violin modes will decrease their frequencies, but
since the specific design of the next generation cryogenic interferometer has still to be defined, we preferred to report just the
overall behavior of the sensitivity. Our analysis remains nonetheless valid in all the experimentally interesting frequency bandwidth, because, due to their
high quality factors, the resonant peaks are relevant in negligible frequency intervals only.

As also shown in Ref.~\cite{qlock2}, one is able to go even below the contribution of the radiation pressure noise
of the control mirror $r$ (curve (g) in Fig.~2) thanks to the optimization of the loop phase $\theta$.
If this phase optimization (which typically involves adding another tunable cavity within the loop)
is not performed, the sensitivity improvement is less evident and, at the optimal feedback gain, the displacement sensitivity tends to coincide
with that determined by the radiation pressure noise of the control cavity (see Ref.~\cite{qlock1}).

\begin{table}
\caption{Parameters used to evaluate the amplitude spectral densities reported in Fig.~\ref{fig2}}
\label{Table:Parameter}
\begin{center}
\begin{tabular}{|l|c|}
\hline
Parameter       &   Value           \\
\hline
Temperature $T$ &     $5$ K \\
$A_0$           &   $2 \pi\cdot 10^{-7} m/Hz^{1 \over  2}$           \\
$\phi_0$        &   $10^{-9}$       \\
$\phi_j (j>0)$    &     $10^{-8}$       \\
$\Omega_0$        &     $2 \pi \cdot 0.6$ Hz \\
$\Omega_1 $     &     $2 \pi \cdot 5649$ Hz \\
$\Omega_j (j>1) $ &     Virgo like \\
$P_{Laser}=2 P_a$   &     $200$ Watt \\
$P_{b}$     &   $200$ Watt \\
${\cal F}_a = {\cal F}_b$     &   $600$ \\
Power Recycling $C_a$   &   $50$ \\
$\eta_a = \eta_b$   &   $0.93$ \\
$L_a$           &   $3000$ m \\
$L_b$           &   $0.1$ m\\
\hline
\end{tabular}
\end{center}
\end{table}

\section{Conclusions}

We have made a careful analysis of the quantum locking scheme proposed by Courty \emph{et al.} \cite{qlock1,qlock2}, based on the application
of an appropriate phase-sensitive feedback loop able to suppress the
radiation pressure noise on cavity mirrors. We have considered in particular its implementation under realistic conditions, i.e., in the presence
of thermal and seismic noise and considering inefficient detectors. We have then considered the application of the quantum locking idea to
a GW interferometer, by assuming to apply an active control to each of the four interferometer mirrors. We have seen that in the case of next generation cryogenic
interferometers, one can achieve an improvement of almost two orders of magnitude in the sensitivity within the $5 - 200$ Hz frequency interval, by applying
quantum locking to each mirror.

\section{Acknowledgements}

This work has been partially supported by INFN through the project ``QuCORP''.

\end{document}